\documentstyle{article}

  \advance\oddsidemargin  by -1in
  \advance\evensidemargin by -1in
  \advance\topmargin  by -1in
\begin{document}

\hfill Pis'ma v Zh.\ Eksp.\ Teor.\ Fiz.\ {\bf 56}, 523 (1992)
[JETP Lett.\ {\bf 56}, 510 (1992)]

\vspace*{1cm}
\begin{center}
{\Large\bf Once again about interchain hopping }
\bigskip\bigskip\bigskip \\
Victor M.~Yakovenko
\bigskip \\
Department of Physics and Astronomy,
Rutgers University, P.\ O.\ Box 849, Piscataway, NJ 08855-0849, USA
\bigskip \\
and
\bigskip \\
L.~D.~Landau Institute for Theoretical Physics, Russian Academy of
Sciences, Kosygin St. 2, Moscow, 117940, Russia
\bigskip\bigskip\bigskip
\end{center}

\begin{center} ABSTRACT \end{center}

   Renormalization group equations and a phase diagram are derived for
a system of two chains with a single-electron hopping between chains
in order to correct the results of a recent paper by F.~V.~Kusmartsev,
A.~Luther, and A.~Nersesyan, Pis'ma v Zh.\ Exp.\ Teor.\ Fiz.\ {\bf 55},
692 (1992) [JETP Lett.\ {\bf 55}, 724 (1992)].

\bigskip\bigskip

   In a recent paper \cite{KLN}, the effect of a single-electron
hopping (SEH) between chains in a system of two chains was studied by
means of the bosonization technique and the Coulomb gas description.
Possible phase transitions of the Berezinskii-Kosterlitz-Thouless
type, which result in generation of pair hoppings between the chains,
were considered.  Unfortunately, despite the spirit of the paper is
correct, Eq.\ (4) turns out to be incomplete and leads to a wrong
phase diagram shown in Fig.\ 1.

   The considered problem is the old one which recently attracted
attention again \cite{All}. The generation of the pair hoppings in a
gapless case was discovered for the first time in Ref.\ \cite{BY} via
a perturbational theory.  Later, these results were confirmed by a
renormalization group (RG) approach \cite{BC}.  The aim of the present
paper is to correct the RG equation (4) and the phase diagram of Ref.\
\cite{KLN} and achieve in this way an agreement with the previously
known results \cite{BY}, \cite{BC}.

   Due to the generation of the pair-hopping terms, the action of the
model,
\begin{equation}
S_0=\int d\tau\,dx \left[\frac{u}{2}((\partial_\tau\Phi)^2+
(\partial_x\Phi)^2)+\frac{2t_\perp}{\pi\alpha}
\cos\frac{1}{2}\gamma\Phi\cos\frac{1}{2}\tilde{\gamma}\tilde{\Phi}
\right],
\label{S0}\end{equation}
has to supplemented with the following term:
\begin{equation}
S_1=\int d\tau\,dx \left[2J\cos\gamma\Phi+2\tilde{J}
\cos\tilde{\gamma}\tilde{\Phi}\right].
\label{S1}
\end{equation}
The notation of Ref.\ \cite{KLN} is used everywhere. In the fermion
representation, the coefficients $J$ and $\tilde{J}$ in Eq.\
(\ref{S1}) are the amplitudes of the electron-electron (EEPH) and the
electron-hole pair hoppings (EHPH), respectively:
\begin{equation}
H_1=(2\pi\alpha)^2\int dx
(J\Psi^+_{1,1}\Psi^+_{2,1}\Psi_{2,-1}\Psi_{1,-1}+
\tilde{J}\Psi^+_{1,1}\Psi^+_{2,-1}\Psi_{2,1}\Psi_{1,-1}+{\rm h.c.}).
\label{H1}
\end{equation}

   The RG equations for the considered system can be derived using the
approach of Ref.\ \cite{W}:
\begin{eqnarray}
&&dt_\perp/dl=(2-0.5K-0.5\tilde{K})t_\perp, \label{dt} \\
&&dJ/dl=2(1-\tilde{K})J+(\tilde{K}-K)t_\perp^2/2\pi v_F,
\label{J1} \\
&&d\tilde{J}/dl=2(1-K)\tilde{J}+(K-\tilde{K})t_\perp^2/2\pi v_F,
\label{J2} \\
&&t_\perp(0)=t_0,\;\;\;J(0)=\tilde{J}(0)=0. \label{0}
\end{eqnarray}
The second terms in the r.h.s.\ of Eq.\ (\ref{J1}) and (\ref{J2})
reflect the generation of the pair hopping terms by combining the SEH
terms. The first terms in the r.h.s.\ of Eq.\ (\ref{dt}) -- (\ref{J2})
reflect the RG dimensions of the appropriate operators. They are
plotted in Fig.\ 1. The renormalization of $K=1/\tilde{K}$ and the
generation of irrelevant terms are neglected. The value $K=1$
corresponds to the case of non-interacting electrons.  In the Hubbard
model, there are certain limitations on the range of possible values
of $K$, however, these limitations do not apply to a generic model,
like an extended Hubbard model \cite{Sch}. Eq.\ (\ref{dt}) --
(\ref{J2}) are essentially the same as Eq.\ (84) of Ref.\ \cite{BC},
although there are some differences in details.

Eq.\ (\ref{dt}) -- (\ref{J2}) with the initial conditions (\ref{0}) have
the following solution (see Eq.\ (87) of Ref.\ \cite{BC}):
\begin{eqnarray}
&&t_\perp=t_0e^{(2-0.5K-0.5\tilde{K})l},
\label{t} \\
&&J=\frac{t_0^2}{2\pi v_F}\frac{\tilde{K}-K}{(2-K-\tilde{K})}
\left(e^{(4-K-\tilde{K})l}-e^{2(1-\tilde{K})l}\right).\label{J}
\end{eqnarray}
The formula for $\tilde{J}$ can be obtained from Eq.\ (\ref{J}) by
exchanging $K$ and $\tilde{K}$.

   Eq.\ (\ref{t}) shows that the SEH is relevant (increases upon
renormalization) if the condition
\begin{equation}
2-0.5K-0.5\tilde{K}>0\;\;\;\Leftrightarrow\;\;\;2-\sqrt{3}<K<2+\sqrt{3}
\label{K1} \end{equation}
is fulfilled. Considering the behavior of the EEPH amplitude $J$, the
two different regimes can be distinguished. If
\begin{equation}
4-K-\tilde{K}>2-2\tilde{K}\;\;\;\Leftrightarrow\;\;\;K<1+\sqrt{2},
\label{K2} \end{equation}
then the first term in Eq.\ (\ref{J}) dominates, and $J(l)$ grows
essentially as $t^2(l)$. If the opposite condition
\begin{equation}
K>1+\sqrt{2}
\label{K3} \end{equation}
is fulfilled, then the second term in Eq.\ (\ref{J}) dominates, and
$J(l)$ grows {\it faster} than $t^2(l)$. Analogously, the EHPH
amplitude $\tilde{J}$ grows faster than $t^2(l)$ if the condition
\begin{equation}
4-K-\tilde{K}<2-2K\;\;\;\Leftrightarrow\;\;\;\tilde{K}>1+\sqrt{2}
\label{K4} \end{equation}
is fulfilled. The mutual position of regions (\ref{K1}) --
(\ref{K4}) is illustrated in Fig.\ 1.

   Upon renormalization, an amplitude $t(l)$, $\alpha J(l)$, or
$\alpha\tilde{J}(l)$ may become of the order of the Fermi energy
$\varepsilon_F\sim v_F/\alpha$, and a crossover to a different
physical regime will take place at the corresponding temperature
$T=\varepsilon_Fe^{-l}$. In the region, labeled as SEH in Fig.\ 1, the
single-electron hopping amplitude $t$ becomes of the order of
$\varepsilon_F$ first of all. In the case of two chains, the
characteristic temperature represents a renormalized energy splitting
of the two chains. In the case of an infinite array of chains, this
temperature marks a crossover from a 1D Luttinger liquid to a 2D or 3D
standard Fermi liquid (FL). There may be phase transitions to ordered
states at lower temperatures.

   On the other hand, in the region marked as EEPH, the
electron-electron pair hopping amplitude $\alpha J(l)$ becomes of the
order of $\varepsilon_F$ when $t(l)$ is still small. In the case of
two chains, the superconducting phases of the two chains become
strongly bound below the corresponding temperature. In the case of an
infinite array of chains, a phase transition to a superconducting (SC)
state takes place at this temperature. It is essential that in the
EEPH region, unlike in the SEH region, the phase transition takes
place directly from the Luttinger liquid regime without an
intermediate FL regime, and the transition is driven by the pair
interchain coupling.  This scenario is reminiscent of one suggested in
Ref.\ \cite{A} for the high $T_c$ superconductors. The same
consideration applies to the EHPH region in Fig.\ 1 where a phase
transition to a density-wave (DW) state is expected.

   The boundaries between the single-electron and the pair hopping
regimes (Eq.\ (\ref{K3}) and (\ref{K4})) were found for the first time
in Ref.\ \cite{BY}. In contradiction with the text of their paper, the
authors of Ref.\ \cite{KLN} neglect Eq.\ (\ref{J1}) and (\ref{J2}),
thus their phase diagram reflects only condition (\ref{K1}). For this
reason, they essentially repeat a wrong conclusion, made in Ref.\
\cite{Wen}, that a Luttinger liquid regime can exist at zero
temperature.

   The author thanks E.~Abrahams for the support of this work via the
NSF Grant No.\ DMR 89-06958 and the Aspen Center for Physics for the
opportunity to have discussions with A.~Luther.

\bigskip\bigskip

\begin{center} FIGURE CAPTIONS \end{center}

   Fig.\ 1. RG dimensions of $t^2$ (curve A): $4-K-\tilde{K}$; $J$
(curve B): $2-2\tilde{K}$; and $\tilde{J}$ (curve C): $2-2K$ as
functions of $K$ (the right horizontal semi-axis) and $\tilde{K}=1/K$
(the left horizontal semi-axis). The difference in a physical behavior
in the three regions, separated by the dashed lines and labeled by the
abbreviations, is explained in the text.

\end{document}